# MARC: Mining Association Rules from datasets by using Clustering models

Shadi AL SHEHABI [1*] and Abdullatif BABA [2]

[1]Computer Engineering department, University of Turkish Aeronautical Association, Ankara, Turkey
[2]Mechatronics Engineering department, University of Turkish Aeronautical Association, Ankara, Turkey
[*]Corresponding author: shadi.alshehabi@ceng.thk.edu.tr

*Abstract* – Association rules are useful to discover relationships, which are mostly hidden, between the different items in large datasets. Symbolic models are the principal tools to extract association rules. This basic technique is time-consuming, and it generates a big number of associated rules. To overcome this drawback, we suggest a new method, called MARC, to extract the more important association rules of two important levels: Type I, and Type II. This approach relies on multi topographic unsupervised neural network model as well as clustering quality measures that evaluate the success of a given numerical classification model to behave as a natural symbolic model.

## I. INTRODUCTION

The incremental behavior of dataset information size represents a major challenge for data analysts because in such cases knowledge discovery becomes more difficult to achieve. On the other hand, the size and variety of data prevent any manual knowledge discovery to be rapidly performed. Therefore, new advanced techniques could be suggested to overcome this problem. Studying these methods is conventionally called data mining or knowledge discovery in database. In this context, knowledge discovery couldn't be considered a direct process to find implicit, unknown, or useful information from large datasets [1][2]. In fact, this mission requires a full analysis of all dimensions of the inspected data. Despite symbolic models are mostly used to perform knowledge discovery, these methods are characterized by remarkable disadvantages. For example, the generating of association rules, that represent a type of knowledge model, is a time-consuming process that leads to producing a large number of redundant association rules. This undesired performance prevents selecting suitable rules from the large available data which is given in numerical descriptive space with many dimensions, which is the problem that may appear with textual data [3]. In order to avoid this problem, unsupervised numerical models (clustering models) are suggested in this paper to extract only the useful knowledge. Clustering methods are effective tools to collect similar data into clusters. They separate dissimilar data from each other. Thus, the different properties will be separated and weak associations between the properties will be ignored.

The IFP-Growth algorithm was suggested in [4] to represent an novel variant of the original FP-Growth algorithm which should be performed in two scans. While in [5] and [6] "multiple minimum support" is used for mining association rules (ARs). A high-utility pattern tree technique was employed in [7] to restore the FP-tree by regarding different measures. A new algorithm FPS that engages the brief conditions into the mining method was proposed by [8]. An MCFP-tree was introduced in [9] to find rule patterns with numerous restrictions. A pressed and organized transaction classification tree was presented in [10]; to construct this tree the dataset should be at first scanned, then all the transactions are compressed inside the tree, internally various patterns with multi-support constraints are created. The algorithm called Adjusting FP-tree was presented in [11] to perform an incremental mining algorithm that exchanges the FP-tree building when the transaction dataset is refreshed. A new tree structure named (compact pattern tree) was introduced in [12] to scan a dataset only one time; this algorithm gives the same outcome as the FP-tree. The CP-tree is built by using a branch sorting approach that gives a dense frequent-descending tree; the built tree shows good success in interactive mining.

The strength of the relationship and the expanded chi-square were combined in [13] to an associative classification algorithm; a quality measure was adopted here to determine the significant ARs. An AR ranking sensitivity measure was proposed in [14] to infer the sensitivity by deciding the uncertainty-increasing factor depending on Bayesian networks. A novel alternative to mining ARs was introduced in [15] by merging lattice and hash tables to improve the performance compared to methods utilizing only hash tables. A regular structure for learning created ARs was suggested in [16] where they combined a data mining algorithm with statistical measurement methods.

Several methods [17][18][19] and [20] used (k-means) to filter large ARs identified by using the Apriori algorithm. Hence, field specialists can build their decision by using a small set of rules. In [18] they presented a structure for mining the relationships of interest. DBSCAN was employed to cluster models from a large dataset of photos in order to produce three sets of patterns: global, local, and categorization.

An algorithm was presented in [20] that divides subsets into identity classes and forms the class graph; where maximal uniform clustering will be introduced to get the most frequent subsets. In [19] they outline how to generate ARs into partitioned hypergraphs algorithm to get the most common clusters.

In [17] they used the DBSCAN algorithm to map transaction data then they applied to their dataset the probabilistic-based algorithm called MPP (Mean-Product of Probabilities). According to [21], the smart search and the optimization algorithms represent the best methods for dealing with complex numerical association rules mining problems. Different methods can solve the problems of symbolic models; for example, an algorithm has been already proposed [3] for extracting simple association rules between a couple of items from a numerical dataset, it still generates a lot of association rules. In [22] the attributes are clustered to reduce the variables count, and then the ARs are extracted from the targeted variables and the earlier built sets of variables.

In this paper, we introduce an alternative algorithm, called MARC, for extracting complex association rules from a numerical dataset of items by means of clusters generated by an unsupervised neural network model called MultiSOM and some clustering quality measures.

## II. MULTISOM CLUSTERING MODEL

MultiSOM model is a neural clustering model that relies on generating several Self-Organizing Maps via the usage of two mechanisms, the generalization mechanism and the inter-communication mechanism [23]. In this paper, we are interested in the generalization mechanism that summarizes the contents of the original SOM map into more general topics.

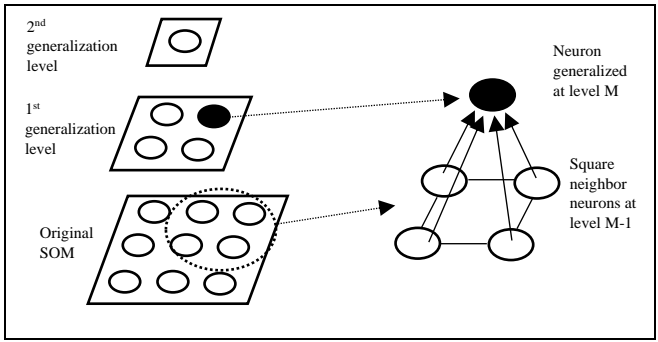

Fig. 1. MultiSOM generalization mechanism

Let i×j be the dimension of the original SOM map (that represents also the number of neurons) then the dimension of the first generalization level will be (i-1) × (j-1), thus the dimension of the kth generalization level will be (i-k) × (j-k) as shown in Fig. 1. The codebook vector of a neuron (cluster) n at the M$^{th}$ level is computed as:

$$W_n^M = \frac{1}{4} \sum_{n_k \in V_n^{M-1}} W_{n_k} \qquad (1)$$

$V_n^{M-1}$ is the square neighbor at level M-1 related to the neuron (n) at the generalization level M. The fact that the generalization mechanism produces homogeneous hierarchical levels, may lead to consider it as a hierarchical clustering method.

## III. MULTI-TOPOGRAPHIC CLUSTERING FOR ASSOCIATION RULES EXTRACTION

Association rules are considered as a type of knowledge extraction from a dataset. The association rules figure out the relationships between the features in a dataset, and they have the following form: P → Q where P and Q are two different sets of features such that P∩Q=∅.

The quality of an association rule is commonly evaluated by two measures, the support (sup.) and the confidence (conf.) [24]. The support of the rule P → Q measures the proportion of the objects that contain P∪Q and it is given as:

$$\sup(P \rightarrow Q) = \frac{\text{support}(P \cup Q)}{N} \qquad (2)$$

Where the support (P∪Q) represents the number of objects that contain both P and Q. N is the number of objects considered in the given dataset.

The confidence of the rule P → Q measures the proportion of objects that contain the features P, as well as the features Q, and it is given as:

$$\text{conf}(P \rightarrow Q) = \frac{\text{support}(P \cup Q)}{\text{support}(P)} \qquad (3)$$

The traditional symbolic methods, such as Apriori [25] and FP-Growth [26], generate all the possible associations between the itemsets and present them by generating a lot of rules. To decrease the number of extracted rules, while keeping the most important ones, two thresholds predefined by the user should be used, the minimum support (minsup) and the minimum confidence (minconf). But this method to select the association rules eliminates a lot of important rare features and keeps a lot of weak associations between the features. To overcome the upper mentioned problems related to symbolic methods, multi topographic unsupervised neural network clustering models such as Neural Gas, Multi-Gas (MGAS), Self-Organizing Map (SOM) and Multi Self-Organizing Map (MultiSOM), have been suggested [3] to extract useful knowledge. MultiSOM contains a generalization mechanism; such that several generalization levels can be generated, at each new high level a lower number of clusters (neurons) are generated at a direct low level.

In this paper, we propose an algorithm, called MARC that relies on the generalization mechanism of MultiSOM to extract association rules. Our proposed algorithm takes into account two clustering quality measures; precision and recall [27]. In this context, the precision measures the homogeneous proportion of the content of the clusters. Thus, the precision of a given feature (t) in a cluster (c) gives the percentage of objects that contains this feature as illustrated in the following equation:

$$\text{Prec}_c(t) = \frac{N_c^t}{|c|} \qquad (4)$$

$N_c^t$ is the number of objects in the cluster (c) that contains the feature (t). |c| is the number of objects in the cluster (c). If $\text{Prec}_c(t) = 1$, then all objects in the cluster (c) contain the feature (t).

The recall gives an idea about the exhaustiveness of the content of the given clusters, in order to evaluate to what extent peculiar items are associated with single clusters. Thus, the recall of a feature (t) in the cluster (c) gives whether the feature is specific and exclusive for the cluster (c) or general and exists in some other clusters. It is given as:

$$\text{Rec}_c(t) = \frac{N_c^t}{N} \quad (5)$$

N is the total number of objects in the dataset.

If $\text{Rec}_c(t) = 1$, then the feature (t) exits only in the cluster (c). For selecting the peculiar features associated with clusters, the feature weights play an important role in eliminating and keeping features of any cluster. The set of peculiar features associated with a cluster (c) is given as:

$$P_c^* = \{t \mid W_c^t > W_{c'}^t\} \quad (6)$$

$$W_c^t = \frac{\sum_{d \in c} w_d^t}{\sum_{c'} \sum_{d \in c'} w_d^t} \quad (7)$$

$w_d^t$ is the weight of the feature (t) in the object (d).

---

**MARC Algorithm:**
1. For all cluster $c \in C$ at a generalization level
2. Find $P_c^*$ the set of peculiar features associated to c
3. Create A (set of features) such that:
   A = { $t_i$ | $t_i \in P_c^*$, Prec ($t_i$)=1, Rec ($t_i$)=1 }
4. Create B (set of features) such that:
   B = { $t_i$ | $t_i \in P_c^*$, Prec ($t_i$)≠1, Rec ($t_i$)=1 }
5. Create *E* (all possible subsets of B) such that:
   E = { $b_k$ | ∃ d∈$D_c$, $b_k \subseteq d$ }
   $b_k$ being a subset of B
   *d* being an object from $D_c$
   $D_c$ being the set of objects associated to a cluster
   $D_c$ = { d | d ∈ c }

   // *Extract association rules of **Type I**:*
6. if |A| ≥ 2 then
7.   For all $t_i \in A$
8.     $t_i \rightarrow A \setminus \{t_i\}$  // ( informative rule )
9.   End for
10. End if
11. if A≠∅ and E≠∅ Then
12.   For all $b_k \in E$
13.     $b_k \rightarrow A$
14.   End for
15. End if

   // *Extract association rules of **Type II**:*
   // extract association rules from the local dataset
   // $D_c$ and its peculiar features $P_c^*$ using Apriori
   // method
16.   Apriori($D_c$, $P_c^*$)
17. End for

Fig. 2. MARC Algorithm for Extracting Complex Association Rules from Clustering Model

Fig. 2 shows our proposed MARC algorithm that extracts association rules from each cluster in a generalization level, where two types of rules could be extracted:

Type I: containing the most important association rules; where the association between the items is very significant. These rules are extracted by means of two quality measurements that are the recall, and the precision. Hence, informative rules, where the premise of the rule (the left-hand side) contains one item, and the conclusion of the rule (the right-hand side) contains a closed itemset. This type extracts exact association rules where their confidence values equal to one.

Type II: containing less important association rules than those of type I. Rather than extracting the association rules from the whole dataset using Apriori algorithm that extracts large numbers of redundant rules, and some other rules with weak associations between their items, the association rules are extracted here from the data objects and their peculiar features in each cluster. The generalization levels allow extracting specific and general rules, such that the more the generalization levels the more general association rule.

Now, let's take a look at the next example where we suppose that a clustering of two clusters is applied to a dataset that contains five objects {$d_1$, $d_2$, $d_3$, $d_4$, $d_5$} and six features {$t_1$, $t_2$, $t_3$, $t_4$, $t_5$, $t_6$}. The frequencies of the features in the objects and the distribution of the objects in the clusters are described in Table 1.

Table 1. Clustered dataset

|  |  | $t_1$ | $t_2$ | $t_3$ | $t_4$ | $t_5$ | $t_6$ |
|---|---|---|---|---|---|---|---|
| **Cluster 1** | $d_1$ | 3 | 6 | 4 | 0 | 1 | 0 |
|  | $d_2$ | 5 | 7 | 6 | 0 | 0 | 2 |
|  | $d_3$ | 3 | 5 | 4 | 3 | 0 | 4 |
| **Cluster 2** | $d_4$ | 0 | 0 | 0 | 0 | 6 | 7 |
|  | $d_5$ | 0 | 0 | 0 | 0 | 6 | 5 |

For a Cluster 1:
$P_{Cluster\ 1}^* = \{t_1, t_2, t_3, t_4\}$
A = {$t_1, t_2, t_3$}
B = {$t_4$}
E = {$t_4$}

Association rules of Type I:
$t_1 \rightarrow \{t_2, t_3\}$, confidence = 100%
$t_2 \rightarrow \{t_1, t_3\}$, confidence = 100%
$t_3 \rightarrow \{t_1, t_2\}$, confidence = 100%
$t_4 \rightarrow \{t_1, t_2, t_3\}$, confidence = 100%

Association rules of Type II:
Association rules are extracted from the local dataset constructed from the objects of Cluster 1, {$d_1$, $d_2$, $d_3$}, and its peculiar features $P_{Cluster\ 1}^*$. The local dataset of Cluster 1 is given Table 2.

Table 2. Local dataset for Cluster 1

| Cluster 1 | t1 | t2 | t3 | t4 |
|---|---|---|---|---|
| $d_1$ | 1 | 1 | 1 | 0 |
| $d_2$ | 1 | 1 | 1 | 0 |
| $d_3$ | 1 | 1 | 1 | 1 |

The association between non-peculiar features in cluster 1 such as $t_5$ and $t_6$ with the other peculiar features is not considered.

## IV. EXPERIMENTAL RESULTS

Our test dataset consists of 1000 patents which are divided into four different subfields: the usage, the advantages, the titles, and the patentees. In our experiment, the "usage" will be only considered. The size of the description space of this dataset is 234. Our objective is to extract complex association rules depending on an optimal number of clusters that are generated by the SOM method. In this experiment, 100 clusters (neurons) are only supposed. Then, the generalization mechanism will be applied. Hence; we have the following specific number of neurons assigned to each generalization level: 81, 64, 49, 36, 25, 16, 9, and 4. The confidence of the association rules is used for evaluating the performance of the MARC method. The higher the confidence value the higher association is. Our experiment consists of extracting association rules from the single "usage" subfield. Both the original MultiSOM map and its generalization maps are employed for extracting the association rules. Table 3 shows the number of extracted association rules from the MultiSOM model after applying MARC algorithm. Two types of rules are extracted; the number of association rules of type II is notably more than the number of association rules of type I. We can notice that the 8th generalization level, with four clusters, contains the biggest number of association rules. This is due to the few numbers of clusters in that level which leads to group more data objects in the clusters.

From fig. 3 we can notice that the confidence of extracted rules of type I is 100% and the quality of these rules is better than those of type II. Moreover, the confidence values of type II rules, in general, decreases through the generalization levels (i.e., the more the generalization level the less the confidence value is).

From fig. 4, we show that the rule length (i.e., the number of the features of the association rule) varies through the different levels for both types I and II. The longest rules are found at level 2 where 24 association rules from type I are generated. From type II, the longest rules are found at level 8 where 7556 association rules are generated but at the expense of their quality. Such that the length of the rules extracted in the first five levels is less than those generated in the 8th level. Fig. 5 shows three different informative association rules extracted using MARC algorithm. All these rules have confidence values equals to 100%. The length of the first is 4; the second is 5 while the third is 3.

Table 3. The number of Complex Association Rules Generated by using MARC Algorithm Table Type Styles

| Generalization levels | Type I | Type II |
|---|---|---|
| Original (100 clusters) | 63 | 618 |
| 1 (81 clusters) | 58 | 90 |
| 2 (64 clusters) | 24 | 620 |
| 3 (49 clusters) | 17 | 216 |
| 4 (36 clusters) | 9 | 728 |
| 5 (25 clusters) | 5 | 768 |
| 6 (16 clusters) | 3 | 340 |
| 7 (9 clusters) | 0 | 390 |
| 8 (4 clusters) | 0 | 7556 |

To show the importance of the proposed algorithm (MARC) we compare it with a well-known symbolic algorithm like Apriori or PF-Growth via the usage of the confidence measure, given that both symbolic methods generates the same amount of association rules. Table 4 illustrates the confidence average and rules length average for the MARC method and the symbolic method such that minsup for the symbolic method is set to 2. MARC method outperforms the symbolic method in terms of confidence for both types I and II. But the symbolic method extracts longer rules than the proposed one because the symbolic method keeps all the possible association, weak and strong, between features whereas MARC method ignores the weak association which leads to extract shorter rules but with higher confidence. Moreover, the association rules with high quality are found in type I and the length average is bigger since this type extracts informative association rules.

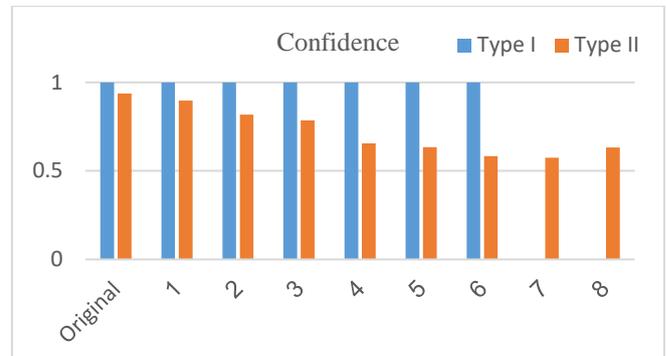

Fig. 3. The confidence average of the extracted rules at different generalization levels by using MARC method

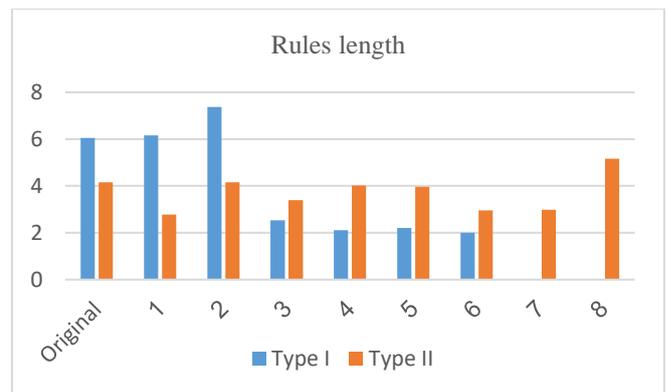

Fig. 4. The length average of the extracted rules at different generalization levels by using MARC method

1. Oils using → sulfur-based load additive, synergistic combination of phosphorus, synthetic oil-based turbo
2. Acid material → combustion chamber, combustion gas, deterioration due, sludge component.
3. Ethylene content ethylene-propylene → lower ethylene content ethylene-propylene, shear stable dispersant.

Fig. 5. Some of complex association rules extracted by using MARC algorithm

Table 4. A comparison between the symbolic methods, AprIorI or FP-Growth, and MARC method

|  | Symbolic method (Apriori or FP-Growth) | MARC algorithm with MultiSOM | |
|---|---|---|---|
|  |  | Type I | Type II |
| Confidence | 0.6262 | 1 | 0.724 |
| Rules length | 5.77 | 4.06 | 3.73 |

V. CONCLUSION

The concept of discovering knowledge has been exposed in this paper from large amounts of data throughout the association analysis of its features. The symbolic methods such as Apriori, FP-Growth, and their variant methods generate a large number of association rules despite the small size of the dataset, this leads to generating a lot of redundant rules from large datasets. If we only use some measures like support and confidence for pruning weak and associations between the items, a lot of important rules will be deleted, especially the rare itemsets that are very important in some numeric datasets like textual datasets. Therefore, the numeric method has been recently adopted, especially neural clustering methods to discover knowledge, such as the MultiSOM method. This method has solved several problems related to symbolic methods. Numeric methods characterized by a great ability, to sum up, data and let the weak associations. As well as, it has a low cost to discover knowledge comparing with symbolic methods. Therefore, we propose here an algorithm, called MARC, that doesn't use any thresholds values for selecting the association rules. This algorithm can extract association rules using the generalization mechanism, provided by an unsupervised neural network method called MultiSOM, with the usage of two measures are; precision and recall, to extract two types of association rules. These measures are used to keep the important features in the clusters as well as extracting the important associations between the itemsets. The obtained results in this paper have shown that the MARC method can overcome all the related problems to the symbolic methods and it can extract complex association rules from numerical datasets.